\documentclass[a4paper]{article}

\usepackage{lineno,hyperref}
\modulolinenumbers[5]

\usepackage{xcolor}
\usepackage{graphicx}

\begin{document}


\title{The first results of monitoring observations of a meteor echo at ISTP SB RAS EKB radar: algorithms, validation, statistics}


\author{Fedorov R.R.
\and
Berngardt O.I.}



\maketitle

\begin{abstract}
The paper presents the implementation of algorithms for automatic
search for signals scattered at meteor tracks based on the ISTP SB
RAS EKB radar data. The algorithm is divided into two phases - the
detection of a meteoric echo and the determination of its parameters.
In general, the algorithm is similar to the algorithms used for specialized
meteor radars, but uses direct fitting of received signal
quadrature components by a model when finding the parameters of meteor
trail. Based on the analysis of maximum of the Geminid meteor shower
13.12.2016 the aspect sensitivity of the scattered signals
detected by the algorithm was shown . This proves that the detected signals correspond
to scattering by irregularities elongated with the direction to the
radiant of the meteor shower. This also confirms that the source of
signals detected by the algorithm are meteor traces. In this paper,
we solve the inverse problem of reconstructing the vector of the neutral
wind velocity from the radar data by the weighted least-squares technique.
A comparison of the roses of horizontal neutral winds obtained in
the model of two-dimensional (horizontal) wind and three-dimensional
wind is made. Their similarity is shown and expected predominance
of the zonal component over the meridional one is shown too. The implemented
algorithm allows us to process the scattered signals in the real time.
The detection algorithm has started its continuous operation at the
ISTP SB RAS EKB radar at the end of 2016.
\end{abstract}



\section{Introduction}
The study of meteor showers is currently one of the tasks of investigating
asteroid-comet hazards. The study of meteor showers and individual
meteors makes it possible to investigate the relationship of meteor
showers with parental bodies (asteroids and comets) and thereby study
the statistics and dynamics of small bodies in the solar system. In
application to the physics of the upper atmosphere, meteors are more
likely the sources of disturbances in the atmosphere, and allow one
to investigate the processes of generation of small-scale irregularities,
as well as their dynamics caused by the background processes in the
upper atmosphere. One of the most common ways to use meteors to
monitor the upper atmosphere is to study the neutral wind velocity.
The burning of meteors occurs at altitudes of about 60-100 km, and
lower, depending on the meteoroid mass \cite{Arecibo_autodetectJGRA:JGRA19862}.
The processes taking place from glow and ionization (for small meteors)
to the electrophone effects and shock waves (for large bolides) are
variable and complex \cite{GRL:GRL25761_janckes_diff_ablation,berngardt_chelybinsk2013},
and are widely studied \cite{Zhu01042016}. The most massive meteoroids
can burn partially, reaching the Earth's surface. Smaller meteors
burn in the atmosphere nearly completely, allowing us to estimate the
mass of the meteoroid based on the burn dynamics.

Radar observations are one of the most common techniques for observing
meteors. In radar observations, two types of scattered signal
are identified based on the scattering object nature: the scattering
by the burning body (head-echo) and scattering by the meteor trail
(trail-echo). 

Scattering on the meteoroid body is more isotropic. The scattering
cross section is smaller than the cross section for scattering on
a meteor trail. Due to this the studies of the head-eacho requires
high potential radars. These radars include MU radar\cite{Kero01092012},
ALTAIR\cite{Close_ALTAIR_headecho}, and some of the incoherent scatter
radars (Millstone-Hill\cite{Millstone-Hill_meteor}, EISCAT\cite{SzaszEISCAT_meteor},
Arecibo\cite{MATHEWS2003_ARECIBOmeteor}). One of the problems
solved by the studies is meteor trajectory investigations, allowing
one to study meteor showers and their sources. In this case, it is
possible to find the trajectory directly, as well as to calculate
the velocity of meteoroid entry into the atmosphere, and to calculate
the deceleration rate of the body\cite{SzaszEISCAT_meteor,MATHEWS2003_ARECIBOmeteor,MathewsARECIBO_meteor}.
Joining these data together allows one to estimate the mass of a
meteoroid and the direction to the radiant.

Another way to find the radiant of meteor shower is the use of trail
scattering and aspect sensitivity of the meteor trail \cite{LovellMeteorAstronomy,CampbellBrown2008144,jonesCMOR}.
The meteor trace is the elongated irregularity of electron density
in the lower ionosphere at heights 60-100 km. Due to this the scattered
signal power is aspect-sensitive and has maximum when radar line-of-sight
is perpendicular to meteor trail (illustrated at Fig.\ref{fig:FIG1}A)\cite{McKinleyMeteorScience}.
Therefore the use of trail scattering requires a relatively low radar
potential and very common in regular studies. 

By the ratio between the plasma frequency in the meteor trail and
the sounding frequency, the meteor trails are traditionally divided
into underdense and overdense ones. Less massive meteoroids correspond
to underdense trails, and more massive meteoroids produce overdense
trails.

Obviously, that less massive meteors prevail over more massive
ones. Therefore the underdense echo is often used in different practical
tasks, for example for studying the dynamics of the upper atmosphere.
The trail formed after the combustion of a meteor is an ionized region
that recombines under the action of the ambipolar diffusion \cite{Jones_and_Jones1990JATP...52..185J}
and the electron density in the trail decreases over time. As
a result, the scattering of the signal at the underdense trail is
characterized by a specific exponential decay of the scattered signal
power. The time of decay is related with the diffusion coefficient,
and this allows researchers to measure this coefficient directly.
In addition, the track is in a dense atmosphere and therefore, is
carried away by a neutral wind at an appropriate altitude due to electron-neutral
collisions. This allows one to measure the dynamics of neutral wind
at the meteor trail altitudes \cite{Nakamura_MUradar7775689,Hall_WIND2006309}.
The use of dense networks of instruments, such as meteor radar networks\cite{deegan1970study}
or SuperDARN radar network\cite{Hall_SDARN_meteors}, allows researchers
to study the motions in the upper atmosphere in a large spatial region
in the monitoring mode.

An important part of meteor studies is the algorithms for automatic
signal detection and for estimating meteor trail parameters. Algorithms
for interpreting the trail-echo usually analyze the power of the received
signal, search for sharp peaks in power with an exponential decay
specific for underdense echo \cite{Tsutsumi_BPMR_autodetect1999}.
Detection of meteor trail echoes at SuperDARN radar is performed by
various algorithms: by studying the average correlation characteristics
of the scattered signal \cite{Jenkins_SDARN_meteor}, by analyzing
the shape of the scattered single-pulse signal with the high frequency
discreetization \cite{RichardToddParrisSDARN_IQ_autodetect}, or by
analyzing the scattered signal from complex sounding sequences \cite{Yukimatu_SDARN_IQ_autodetect}.

In this paper we describe the implementation of algorithms for automatic
search for meteors at ISTP SB RAS EKB radar. Also we validate the
algorithm by using the aspect sensitivity of scattered signals. We
present the first statistics of meteor trails observations at the
radar. Also we describe a method for solving the inverse problem of
reconstructing the vector of the neutral wind velocity from the obtained
data without elevation angle information.

\section{EKB ISTP SB RAS radar}

\begin{figure}
\includegraphics[scale=0.3]{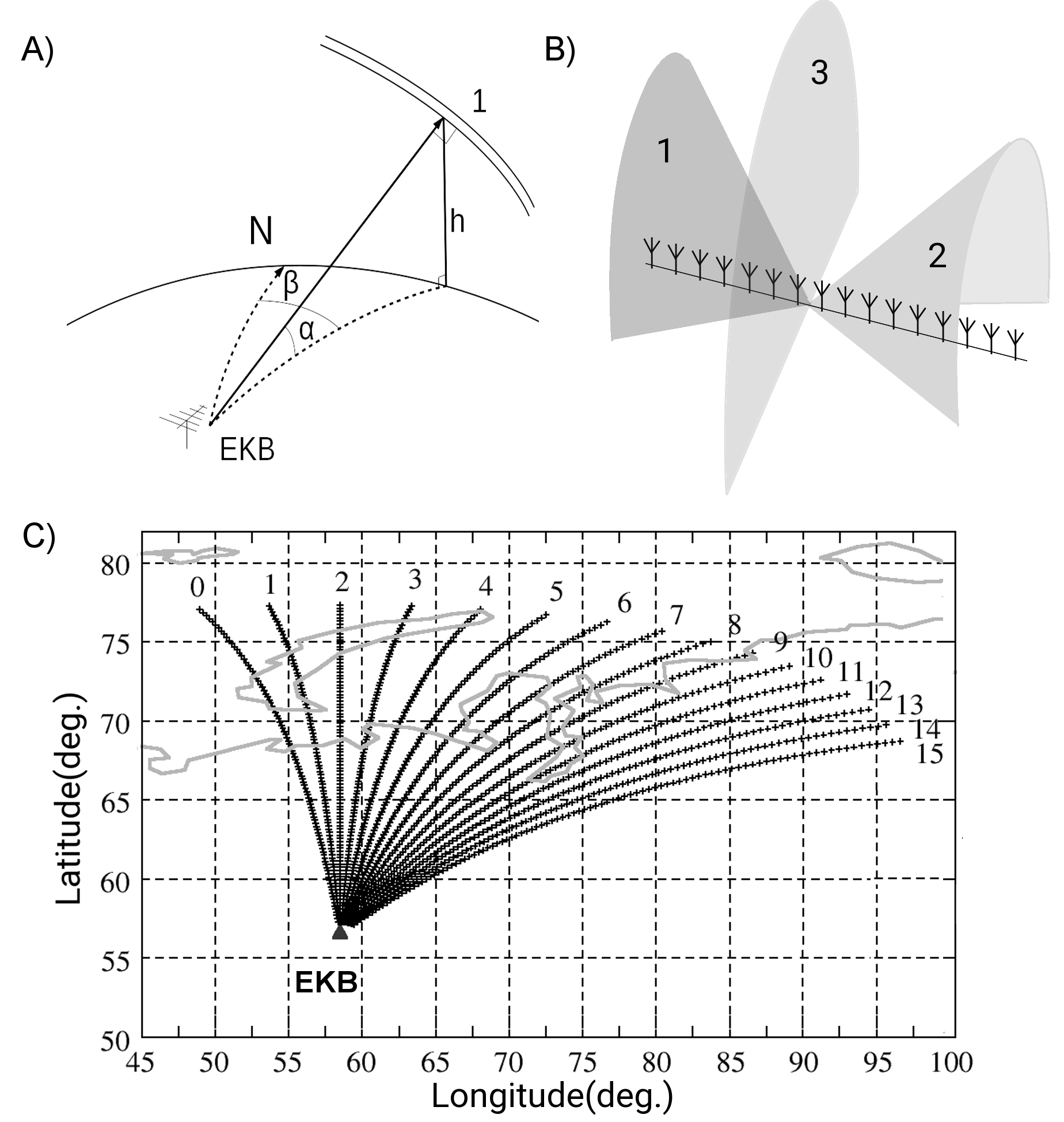}
\caption{A) - scheme of scattering at the meteor trail (1). B) - antenna pattern
surfaces for EKB radar: (1) and (2) - for the cases of outer beams
(beams 0 and 15), (3) - for the case of central beams (beams 7 and
8); C) - field of view of EKB radar.}
\label{fig:FIG1}
\end{figure}

The EKB ISTP SB RAS radar is a decameter over-the-horizon radar, developed
and manufactured at the University of Leicester (UK) and similar to
the SuperDARN radars. The radar is located in the Sverdlovsk Region
of the Russian Federation (56.5 N, 58.5 E). Its field of view has
an azimuthal width of about $52^{0}$ and is divided into 16 directions
(beams) $3^{0}-6^{0}$ width each depending on the sounding frequency.
The central direction of the radar filed of view has the azimuth $19^{0}$
from the North. The radar field-of-view and its beam configuration
are shown in Fig.\ref{fig:FIG1}B,C. The radar operates in 8-20MHz
frequency range, providing a range resolution in standard modes of
15-45 km and has the maximal range about 3000-4500 km. The EKB radar
is a stereo radar of CUTLASS kind and can operate at two channels
(with different carrier frequencies and different beams) simultaneously.
The antenna array of the EKB radar has a symmetry axis, antenna pattern
has a significant back lobe in the regular frequency range of 10-12
MHz, so the surface of antenna pattern for each beam can be considered
as the surface of a cone. The shapes of the antenna pattern at different
beams are shown in Fig.\ref{fig:FIG1}B (according to \cite{Berngardt_2017}).

The standard operational mode of the radar is the transmitting of
various multipulse (Golomb) sequences \cite{Berngardt_Golomb_ruler},
which provide high spatial and spectral resolution simultaneously.
Standard autocorrelation processing of the signal and the accumulation
of the autocorrelation function over the 4-8 seconds for each selected
beam, as well as the subsequent analysis of the amplitude-phase structure
of the averaged autocorrelation function by algorithms specially developed
by SuperDARN community\cite{Ribeiro_FitACF_description}, makes it
possible to estimate the average characteristics of scattered ionospheric
signals, as well as signals scattered from the Earth's surface.

The most common way of processing the meteor echo at SuperDARN radars
is the analysis of the average autocorrelation functions: power, range,
line-of-sight Doppler velocity and spectral width. The meteor trail
detection algorithms are based on the statistical thresholds of these
parameters determined experimentally in joint measurements with meteor
radars. The time resolution of these techniques corresponds to the
accumulation time of the autocorrelation function and is usually about
4-8 seconds \cite{angeo-21-2073-2003}.

\section{Algorithm of meteor trail detection and estimation of its parameters}

\subsection{Basic description of the algorithm}

As it was stated earlier, standard measurements by the radar are carried
out with complex multi-pulse sounding sequences. This allows us to
treat the radar as a meteor radar with 0.3 msec sounding pulse and
with an irregular pulse repetition rate. Time resolution of the radar
changes from 2.4 msec (minimal interpulse interval) to about 36 msec
(maximal interpulse interval). The registration of the full shape
scattered signal (quadrature components) is started after the beginning
of each sounding sequence with standard temporal discretization rate 0.3 msec,
which provides 45 km range resolution of the measurements.

The optimal way of analysing meteor trail scattering is to search
for and to determine the parameters of the scattered signal simultaneously
to improve the algorithm accuracy. However, the need of the real-time
data processing, demanding computer resources and limited computing
resources at the radar did not allow us to implement the most accurate
technique for trail detection and estimation of its parameters. Therefore
the problem was divided into two tasks, solved separately: task
1 - meteor trail detection and estimating of the distance to the scatterer;
task 2 - fitting the model of the trail scattered signal to the data
and estimation of the its parameters. The meteor trail detection was
carried out by the amplitude of the scattered signal, and model fit
and estimation of its parameters was carried out by the analysis of full
structure of the scattered signal (its quadrature components).

\subsection{Task 1 - meteor trail detection}

The task of the detection stage is to search for bursts of signal
amplitude that are potentially suitable for interpretation as a meteor
trail scatter. The radar registers the quadrature components of the
scattered signal. After transfer to zero carrier frequency (heterodyning)
it can be described as a complex signal $u(t)$:

\begin{equation}
u(t)=I(t)+iQ(t)=A(t)e^{i\varphi(t)}\label{eq:signal_iq}
\end{equation}
where $I(t),Q(t)$- quadrature components of the scattered signal,
$A(t)$ - absolute value of the signal amplitude ('amplitude'), $\varphi(t)$
- signal phase.

Since the radar transmits a repeating complex pulse sequence, it is
convenient to represent the amplitude of the scattered signal $A(t)=|u(t)|$
as a function of two arguments - the transmit moment of the nearest
pulse $T_{m}^{e}$ and the radar delay $R$. Here and after the bottom
index $m=0,1...$ denotes the index number of the pulse; the top index
$f$ ('first') marks the response corresponding to the first pulse
of the sequence, the top index $e$ marks the response from corresponding
to any pulse in the sounding sequence - thus $T_{m}^{f}=T_{m}^{e}$
when $m=s\cdot N_{p}$. Here $s=0,1,...$; $N_{p}$ is the number
of pulses in the sequence, value of $T_{m}^{f}$ is not defined for
$m\neq sN_{p}$. These arguments are related to the moment of measurement
of the signal $t$ by the following equation:

\begin{equation}
t=T_{m}^{e}+2R/c\label{eq:time_definition}
\end{equation}

Thus, we proceed from analyzing the amplitude modulus of the signal
$A(t)$ and its quadrature components $I(t),Q(t)$ as a functions
of absolute time, to the analysis of two-dimensional arrays $A(R,T_{m}^{e})$
and $I(R,T_{m}^{e}),Q(R,T_{m}^{e})$ respectively. It should be noted
that due to the peculiarities of the transmitted sounding signal,
which is a non-equidistant sequence of short pulses, the sequence
$\{T_{m}^{e}\}$ is also not equidistant.

In order for the signal to be accepted as a candidate for meteor trail
scattering, the signal must satisfy the following conditions (verification
stages):

\paragraph{Stage 1. High amplitude of the burst}

Meteor trails are relatively rare, relatively shortlived and nearly
stationary objects. We are interested in bursts of high amplitude,
by which we can surely determine the other parameters of the meteor
trail echo: its lifetime and the average line-of-sight velocity. Therefore,
we search for an excess of the amplitude above the noise level. Since
the meteor trails are quasistationary - within the lifetime of the
trail their position (range) $R$ does not depend on the time $T_{m}^{e}$,
the meteor trail search can be carried out at each fixed range $R$
independently. The search condition of the trail is a short-term amplitude
exceeding of some threshold level. When searching, for each radar
range $R_{k}$ the threshold value $M_{k}$ is calculated over the
set of sounding runs $N$ (one session corresponds to one sounding
sequence). Then we select all the coordinates $(R_{k},T_{m}^{f})$
of signals scattered from the first pulses of the sounding sequences,
for which the signal amplitude exceeds the threshold level:

\begin{equation}
A(R_{k},T_{m}^{f})>M_{k}\label{eq:stage1}
\end{equation}

Here $M_{k}$ is the threshold level, defined as the average amplitude
of signals scattered from the first pulses of the sounding sequences,
averaged over the sounding runs:

\begin{equation}
M_{k}=\frac{1}{N}\sum_{m=1}^{N}{\scriptstyle A(R_{k},T_{mN_{p}}^{f})}\label{eq:stage1_border}
\end{equation}

All the solutions $(R_{k},T_{m}^{f})$, satisfying (\ref{eq:stage1}),
are bursts of sufficiently high amplitude and pass to the next stage
of analysis.

\paragraph{Stage 2. High spatial localization.}

Meteor trails are rare and highly localized spatial objects. Therefore,
for each burst of the amplitude found at stage 1, we verify the fulfillment
of the second condition-the spatial localization of the amplitude
burst. For this, the average level at the ranges $R_{k-1},R_{k+1}$
around the trail range $R_{k}$ should not exceed the threshold value
$m_{k}$:

\begin{equation}
\frac{A(R_{k+1},T_{m}^{f})+A(R_{k-1},T_{m}^{f})}{2}<m_{k}\label{eq:stage2}
\end{equation}

As the threshold value $m_{k}$, we use the median value of the amplitude
of the signals scattered from each pulse of the sounding sequence:

\begin{equation}
m_{k}:\{F_{A}(m_{k})=\frac{1}{2}\}\label{eq:stage2_border}
\end{equation}

where $F_{A}(A(R_{k},T_{m}^{e}))$ is an empirical distribution function
of the amplitudes $A(R_{k},T_{m}^{e})$ for a fixed range $R_{k}$,
$m\in[1..K]$, where $K$ is the total number of pulses transmitted
during the scan time.

All the detected bursts with coordinates $(R_{k},T_{m}^{f})$, satisfying
(\ref{eq:stage2}), are spatially localized bursts of amplitude and
pass to the next stage of analysis.

\paragraph{Step 3. Monotonic decrease of signal amplitude}

As already mentioned, the main feature of scattering by underdense
trails is the exponential decay of the signal amplitude over the time.
Therefore, when detecting a meteor trail echo by pulse sequences,
we look for a series of bursts with an amplitude monotonically decreasing
in time. To do this, the sequences of signals scattered by the first
pulse of the probing sequence $T_{m+lN_{p}}^{f}$ are analyzed. They
should follow the pairs $(R_{k},T_{m}^{f})$ detected at the previous
stage. The sequences should satisfy the condition (\ref{eq:stage1})
and should decrease in time (fulfillment of the condition (\ref{eq:stage2})
is not required in this case):

\begin{equation}
A(R_{k},T_{m+lN_{p}}^{f})>A(R_{k},T_{m+(l+1)N_{p}}^{f}),l=0...s-1\label{eq:stage3}
\end{equation}
and form the sequence $S_{k,m,L}$ of signals scattered from each
sounding pulse (hereinafter, 'the set'). The length $L$ of the sequence
is $L=s\cdot N_{p}$ elements, where $s$ is the number of sounding
sequences included in the set $S_{k,m,L}$:

\begin{equation}
S_{k,m,L}=((R_{k},T_{m}^{e}),(R_{k},T_{m+1}^{e}),(R_{k},T_{m+2}^{e})...(R_{k},T_{m+L}^{e}))\label{eq:Seq_def}
\end{equation}

The formation of the set is completed when the fulfillment is terminated
of the condition (\ref{eq:stage1}):

\begin{equation}
\left\{ \begin{array}{l}
A(R_{k},T_{m+sN_{p}}^{f})>M_{k}\\
A(R_{k},T_{m+(s+1)N_{p}}^{f})<M_{k}
\end{array}\right.\label{eq:seq_len1}
\end{equation}
or the condition (\ref{eq:stage3}):

\begin{equation}
\left\{ \begin{array}{l}
A(R_{k},T_{m+(s-1)N_{p}}^{f})>A(R_{k},T_{m+sN_{p}}^{f})\\
A(R_{k},T_{m+sN_{p}}^{f})<A(R_{k},T_{m+(s+1)N_{p}}^{f})
\end{array}\right.\label{eq:seq_len2}
\end{equation}

In this approach, several sets $S_{k,m,L}$ can be generated, and
each of them is processed separately later. The first point $(R_{k},T_{m}^{f})$
of each set $S_{k,m,L}$ corresponding to the scattering of the first
sounding pulse in the first sounding sequence in the set is considered
to be the initial coordinates (range and initial moment) of the meteor
trail. The length $L$ of the sequence corresponds to the duration
of the meteor echo found. In the current implementation of the algorithm,
the same sounding sequence response can be included in several sets.

To reduce the noise in inversion stage, the scattering responses that
do not satisfy the following condition are excluded from the resulting
set $S_{k,m,L}$:

\begin{equation}
A(R_{k},T_{m}^{e})>m_{k}\label{eq:stage4}
\end{equation}
, the total length of the set $L$ is changed accordingly.

\begin{figure}
\includegraphics[scale=0.07]{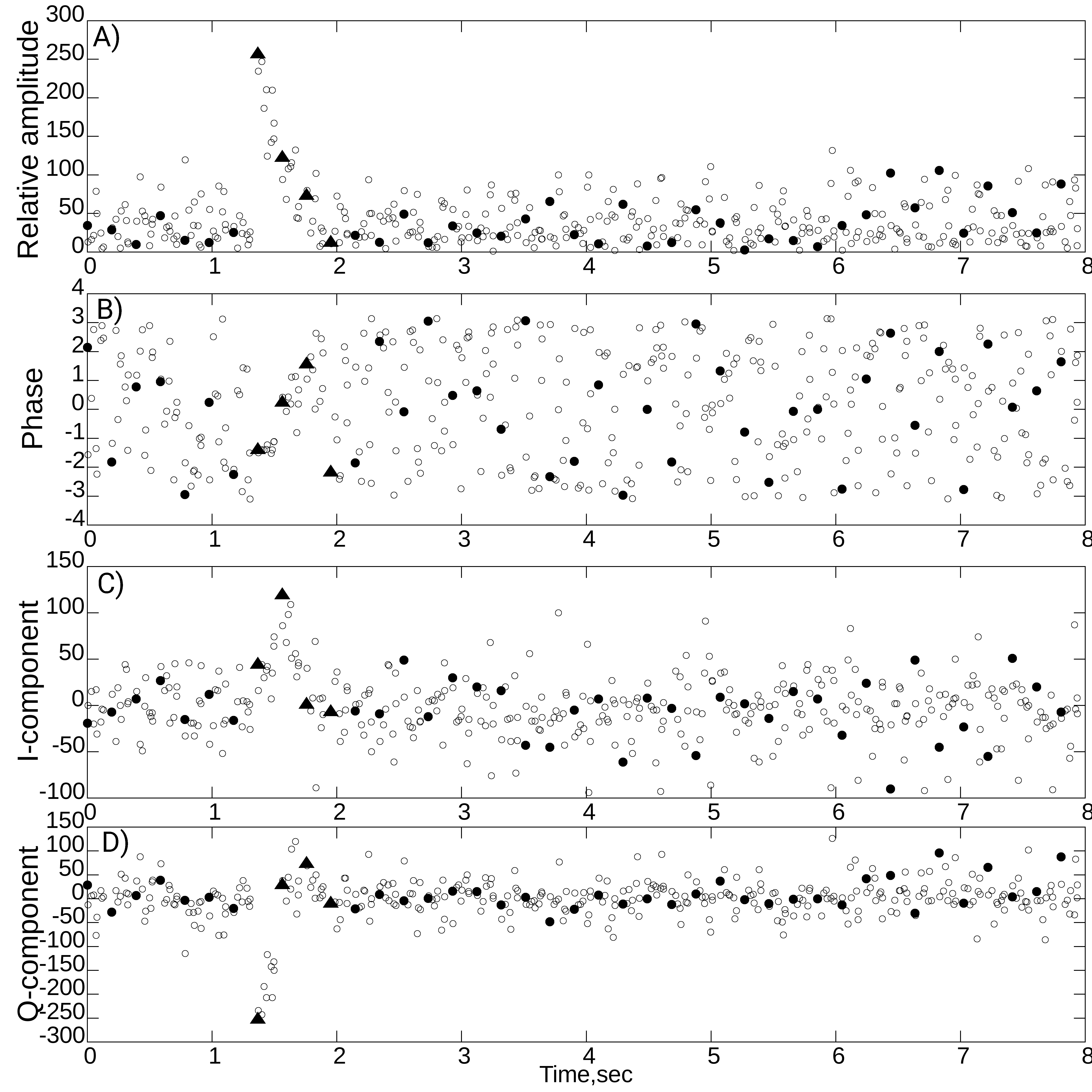}
\caption{Example of the amplitude (A), phase (B) and quadrature components
(C, D) of the signal at a fixed range $R_{k}$ in the presence of
meteor trail scatter. Black triangles denote the first pulses of the
sounding sequences identified in the framework of stage 1, black circles
- the first pulses of the remaining sequences within the scanning
cycle, circles denote the signal from all other sounding pulses.}
\label{fig:FIG2}
\end{figure}

In Fig.\ref{fig:FIG2} is shown an example
of a selected sequence corresponding to a meteor trail echo at a fixed
radar range $R_{k}$ as a function of time $T_{m+i}^{e}$.

\subsection{Task 2 - estimation of meteor trail parameters}

The main characteristics of underdense scattering at meteor trails
are: the coordinates of the combustion point, the characteristic decay
time and the line-of-sight Doppler velocity. For studying the diffusion
processes in the upper atmosphere and the fine structure of the atmospheric
wind it is important to know the combustion height and other spatial
coordinates. EKB radar does not have the ability to determine the
elevation angle for now, and therefore the height can not be determined
from experimental data. 

The amplitude of the signals scattered at underdense trails is monotonically
decayed. The characteristic decay time is related with the recombination
of the trail under the action of ambipolar diffusion and can be used
for estimating the diffusion coefficient \cite{Jones_and_Jones1990JATP...52..185J}.
The meteor burns in dense layers of the atmosphere and therefore the
trail moves under the influence of wind at the altitudes of combustion.
Since the speed and direction of the wind can not change significantly
during the existence of the echo, the Doppler shift of the phase of
the signal is linear in the first approximation. This allows us to measure
the wind line-of-sight velocity \cite{Tsutsumi_BPMR_autodetect1999}
from the phase changes of the scattered signals. 

In our algorithm each trail echo candidate, represented as set (\ref{eq:Seq_def})
with initial coordinates $(R_{k},T_{m}^{f})$
and duration $L$, and satisfying the requirements (\ref{eq:stage1},\ref{eq:stage2},\ref{eq:stage3})
is used to solve the inverse problem of estimating the meteor trail
characteristics.

For solving the inverse problem, we analyse the sequence of scattered
signal quadrature components $I(t_{i}),Q(t_{i})$, where $t_{i}$
is determined by the moment $T_{m}^{e}$ of the meteor
echo start and by the distance $R_{k}$ to it:

\begin{equation}
t_{i}=T_{m+i}^{e}+2R_{k}/c;\,i=0..L\label{eq:def_t}
\end{equation}

The sequence of quadrature components $\{I(t_{i}),Q(t_{i})\}_{i=0..L}$
is used to solve the inverse problem by searching for the parameters
of the model signal that fits it best. An example of the experimental
sequence is shown in the Fig.\ref{fig:FIG2}C-D.

As a model signal, the well-known model \cite{RichardToddParrisSDARN_IQ_autodetect}
is used:

\begin{equation}
\left\{ \begin{array}{l}
u_{m}(t;A_{0},\omega,\tau)=A_{0}(I_{m}(t;\omega,\tau)+iQ_{m}(t;\omega,\tau))\\
I_{m}(t;\omega,\tau)=\theta(t)e^{-t/\tau}cos(\omega t)\\
Q_{m}(t;\omega,\tau)=\theta(t)e^{-t/\tau}sin(\omega t)
\end{array}\right.\label{eq:ModelForFit}
\end{equation}

Here $A_{0}=A_{0I}+iA_{0Q}$ is the initial complex-valued amplitude
of the scattered signal with taking into account the initial phase;
$I_{m}(t;\omega,\tau),Q_{m}(t;\omega,\tau)$ are the real and imaginary
components of the model function as a functions of the time $t$ for
given values of the model parameters $\omega,\tau$; $\theta(t)$
is the Heaviside function (a single step function). The optimal parameters
of the model that should be found are: $A_{0}$ - real and imaginary
components of the initial amplitude of the scattered signal, $\omega$
and $\tau$ - phase slope (cyclic frequency) and decay time, respectively.

The search for the optimal model parameters is carried out by the
least squares method, and the optimal parameters $(A_{0},\omega,\tau)$
should provide the absolute minimum of the residual function $\Omega(A_{0},\omega,\tau)$:

\begin{equation}
\Omega(A_{0},\omega,\tau)=\sum_{i=0}^{L}|u(t_{i})-u_{m}(t_{i}-t_{0};A_{0},\omega,\tau)|^{2}=min\label{eq:OptConditionFunc}
\end{equation}
Here, $u(t)$ is the complex-valued representation of the scattered
signal (\ref{eq:signal_iq}), $t_{i}$ are the moments of meteor trail
scatter observations for given meteor trail; $t_{0}$ is the initial
moment of meteor trail scatter observation for given meteor trail.

The minimization problem (\ref{eq:OptConditionFunc}) is linear over
the parameters $A_{0I},A_{0Q}$ (the parts of complex-valued parameter
$A_{0}$) and is non-linear over the parameters $\omega,\tau$. Therefore
to find the minimum (\ref{eq:OptConditionFunc}) we use a direct search
of parameters $\omega,\tau$ over a grid of values and an analytical
calculation of the optimal parameters $A_{0I},A_{0Q}$ at each point
of the grid.

The following parameter grid is used in the search:

- for the Doppler frequency shift $\omega$ we use equidistant grid
corresponding to Doppler velocities from -200 to 200 m/s in steps
of 1 m/s. In the calculations, the Doppler velocity is converted to
the Doppler frequency shift according to the classical formula $\omega=\frac{4\pi V}{\lambda}$,
where $\lambda$-radar wavelength, $V$ - Doppler velocity. Our choice
of the limits for Doppler velocities is caused by the maximal drift
velocities observed experimentally, and does not exceed the acoustic
speed. The grid step is chosen 10 times more detailed than the standard
velocity resolution of this radar (10 m/s) \cite{angeo-22-459-2004};

- for the decay time we use an equidistant grid from 0.1 to 40 seconds
in steps of 0.1 second. Selection of the step is related with the
duration of the sounding sequence. The maximal grid value has been
selected experimentally: on the one hand, based on the reasons of
calculation speed, required for real time processing, on the other
hand, to determine the lifetime of the most part of the observed meteor
trails.

The optimal amplitude $A_{0}$ at each point of the grid $\omega,\tau$
is defined as:

\[
A_{0I}=\frac{\sum_{i=0}^{L}\left(I_{m}(t_{i}-t_{0};\omega,\tau)I_{e}(t_{i})+Q_{m}(t_{i}-t_{0};\omega,\tau)Q_{e}(t_{i})\right)}{\sum_{i=0}^{L}\left(I_{m}^{2}(t_{i}-t_{0};\omega,\tau)+Q_{m}^{2}(t_{i}-t_{0};\omega,\tau)\right)}
\]

\[
A_{0Q}=\frac{\sum_{i=0}^{L}\left(I_{m}(t_{i}-t_{0};\omega,\tau)Q_{e}(t_{i})-Q_{m}(t_{i}-t_{0};\omega,\tau)I_{e}(t_{i})\right)}{\sum_{i=0}^{L}\left(I_{m}^{2}(t_{i}-t_{0};\omega,\tau)+Q_{m}^{2}(t_{i}-t_{0};\omega,\tau)\right)}
\]
where $t_{0}=T_{m}^{f}+2R_{k}/c$ is the initial moment
of a meteor trail echo observations according to (\ref{eq:def_t}).

The optimal values of the parameters $(A_{0I},A_{0Q},\omega,\tau)$
that provide the minimum of the functional (\ref{eq:OptConditionFunc})
on the grid $\omega,\tau$ are considered by us the actual parameters
of the meteor trail - the complex-valued amplitude $A_{0}=A_{0I}+iA_{0Q}$,
Doppler frequency shift $\omega$ and trail lifetime $\tau$.

\section{Testing the algorithm}

\subsection{Accuracy of the signal model and the inversion technique}

When retrieving the meteor trail parameters described in the previous
section, the accuracy of the model and inversion technique was not
taken into account. To estimate the accuracy of the model parameters,
we checked the accuracy of the inversion technique based on the experimental
data. For this purpose, the distribution of the error between the
model and experimental data was analyzed in terms of amplitude and
phase errors separately based on EKB data during 2016.

The normalized error of the found solution in terms of the amplitude
was determined from the standard deviation between the amplitude of
the experimental data and the model one normalized to the maximal
amplitude of the scattered signal on the meteor trail, according to
the expression:

\[
\sigma_{A,r}(\omega,\tau)=\frac{\sigma_{A}(\omega,\tau)}{|u(t_{0})|}
\]

where
\[
\sigma_{A}^{2}(\omega,\tau)=\frac{1}{L}\sum_{i=0}^{L}\left(|u(t_{i})|-|u_{m}(t_{i};A_{0},\omega,\tau)|\right)^{2}
\]

-the mean-square deviation of the experimental amplitude from its
model value.

The phase error of the solution was determined from the standard deviation
between the experimental phase and the model one:

\[
\sigma_{\varphi}^{2}(\omega,\tau)=\frac{1}{L}\sum_{i=0}^{L}\left(arg\left(u(t_{i})\right)-arg\left(u_{m}(t_{i};A_{0},\omega,\tau)\right)\right)^{2}
\]

where $arg(u)$ is the phase of the complex number $u$.

\begin{figure}
\includegraphics[scale=0.2]{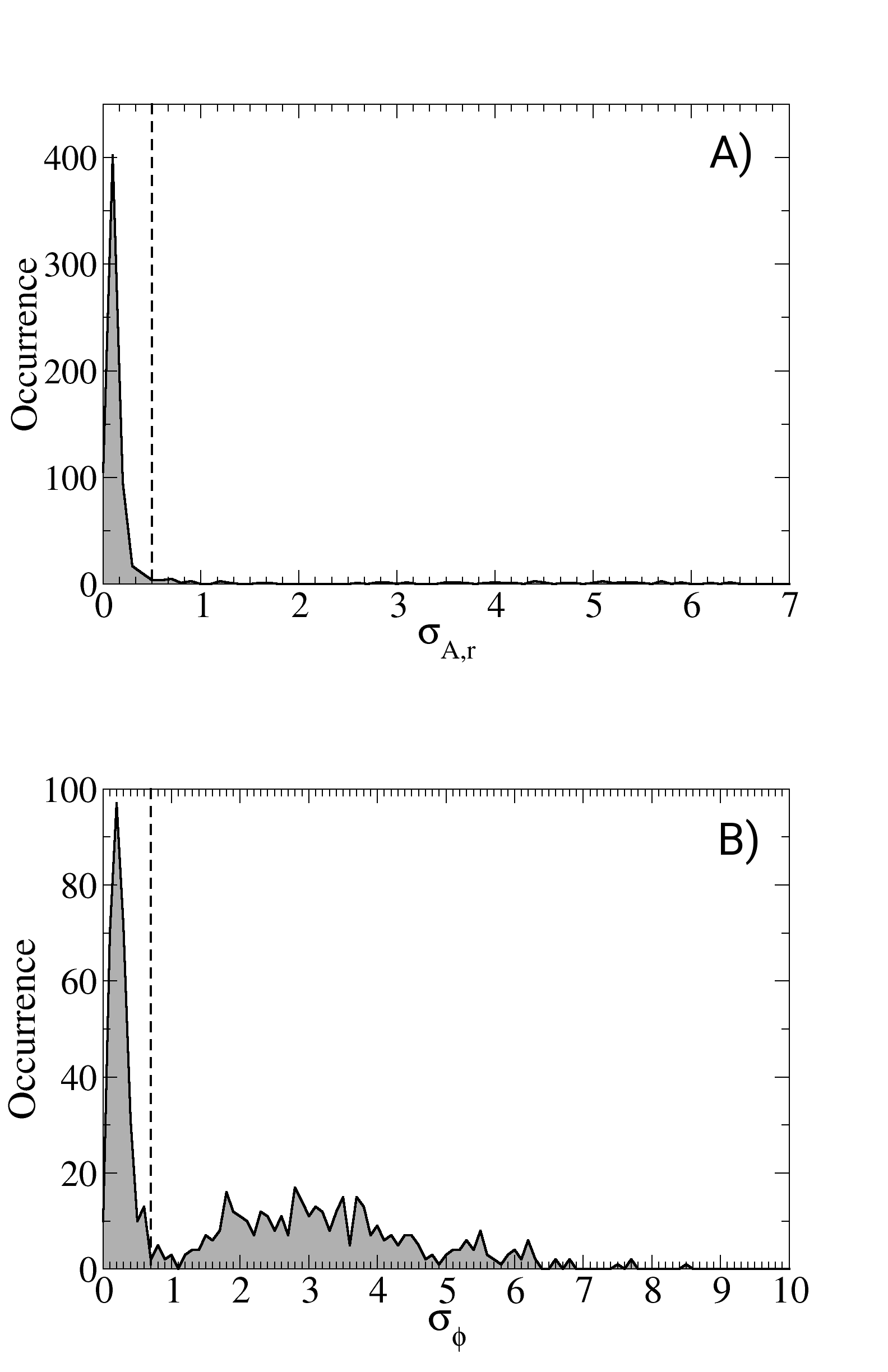}
\caption{Distributions of the errors between experimental
and model values. A) normalized error of amplitude estimation $\sigma_{A,r}(\omega,\tau)$,
B) error of the phase estimation $\sigma_{\varphi}(\omega,\tau)$.
The dashed lines correspond to the threshold values.}
\label{fig:FIG3}
\end{figure}

The distributions of $\sigma_{A,r}(\omega,\tau)$ and $\sigma_{\varphi}(\omega,\tau)$
are shown in Fig.\ref{fig:FIG3}. They
were calculated over the data obtained at EKB radar during maximums
of 6 meteor showers from January to August 2016 (totally 16 days).

Fig.\ref{fig:FIG3} shows the following
two qualitative threshold levels used by us for the data verification:

\begin{equation}
\left\{ \begin{array}{l}
\sigma_{A,r}(\omega,\tau)<0.5\\
\sigma_{\varphi}(\omega,\tau)<0.7
\end{array}\right.\label{eq:sigma_cond}
\end{equation}

Within the threshold level for $\sigma_{A,r}(\omega,\tau)$
the most part of the errors is contained. The threshold level for $\sigma_{\varphi}(\omega,\tau)$
is chosen based on the reasons of an allowable error in determining
the Doppler velocity: the threshold level of 0.7 radians/sec corresponds
to the line-of-sight drift velocity about 2 m/s for 8MHz frequency.
Fig.\ref{fig:FIG3}B
shows that the phase threshold level significantly reduces (by 57\%)
the number of signals that can be interpreted well.

As qualitative analysis shown, the use of the threshold levels allows
us to increase the degree of confidence of the data obtained (by eliminating
the cases when the model of scattering on a single underdense trail
is inadequate to experimental data).

\begin{figure}
\includegraphics[scale=0.07]{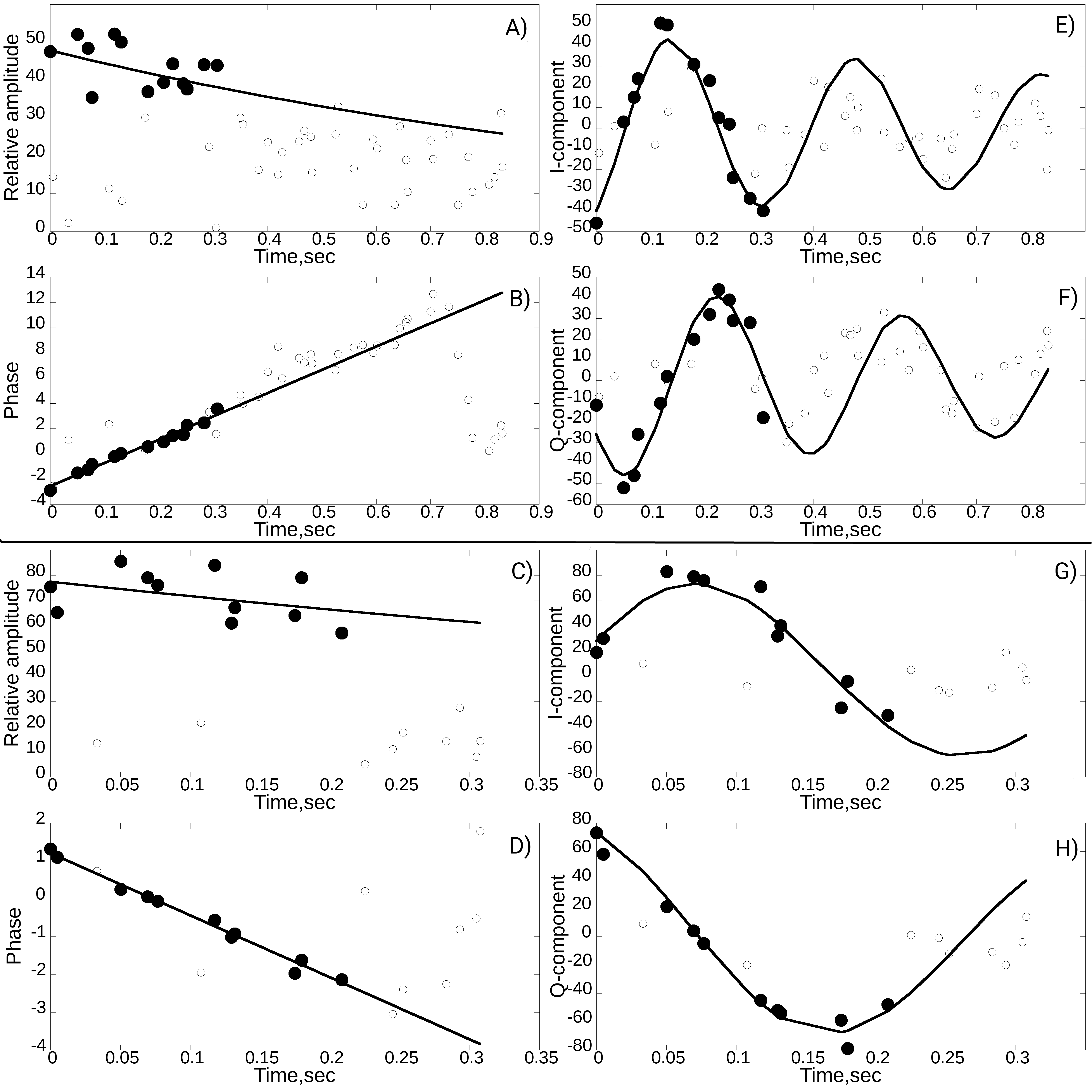}
\caption{Two examples of well-fitted meteor trail echo. A),
C) are the amplitude modulus of the signal $|u(t_{i})|$;
B), D) are the phases of the signal $arg(u(t_{i}))$;
E-H) - I/Q signal components. The black circles are the experimental
signal, the empty circles are the noisy values. The lines correspond
to the model approximation.}
\label{fig:FIG4}
\end{figure}

Examples of records that satisfy both the conditions (\ref{eq:stage1},\ref{eq:stage2},\ref{eq:stage3})
and the quality fit condition (\ref{eq:sigma_cond}) are shown in
Fig.\ref{fig:FIG4}. In panels A) and C) are shown the
amplitudes of the signal, in panel B) and D) are shown the phases
of the signal; in panels E)-H) are shown the quadrature components
of the scattered signal. The black circles in the figure represent
the signal values used for fitting, the empty circles show the points
that are rejected as noise by the criterion (\ref{eq:stage4}). The
lines show the model shape with a minimal deviation of the model from
the experimental data (\ref{eq:OptConditionFunc}).

\subsection{Meteor trail statistics according to EKB radar data}

\begin{figure}
\includegraphics[scale=0.5]{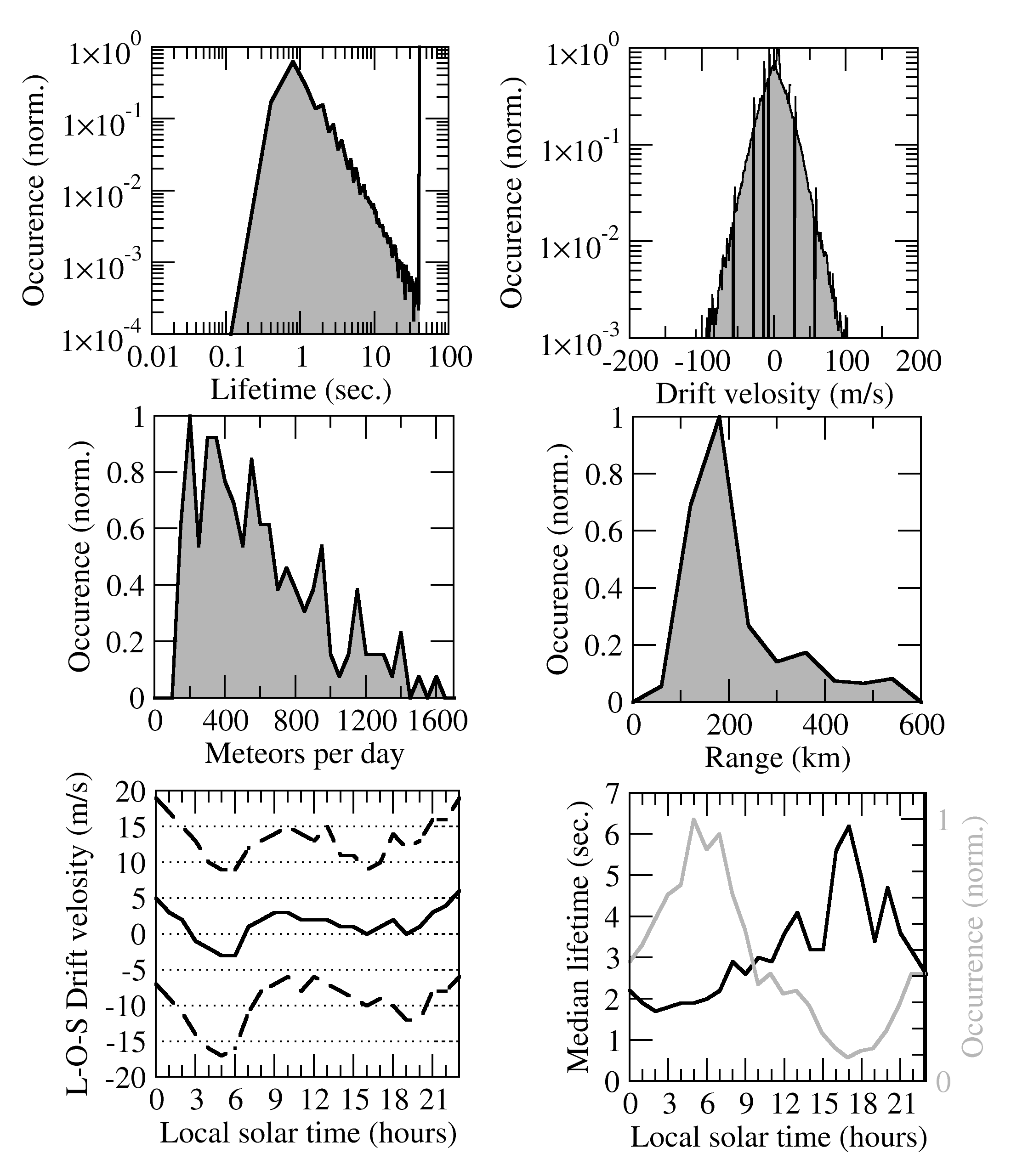}
\caption{Statistics of meteor trail observations at EKB radar from December
2016 to August 2017. A) - meteor trails number as function of their
lifetime B) - the meteor trails number as function of their line-of-sight
Doppler velocity; C) - distribution of the daily number of meteor
trails; D) - distribution of the meteor trails as a function of radar
range; E) - median line-of-sight velocity (black line) and its first
and third quartiles (green lines), as a function of local solar time;
F) - the median lifetime of the meteor trail as a function of local
solar time(black line), 
normalized distribution of meteor trails as a function of local solar
time.(grey line)
}
\label{fig:FIG5}
\end{figure}

Fig.\ref{fig:FIG5} shows the statistics of meteor trail observations
at EKB radar from December 2016 to August 2017. The statistics includes
observations of more than 128 thousand meteors. Fig.\ref{fig:FIG5}A
shows the distribution of the meteor trails over their lifetime. From
Fig.\ref{fig:FIG5} one can see that the most probable lifetime is
about 0.6 seconds, which confirms the correctness of the selected
limits for estimating it. The distribution spire at 40 seconds corresponds
to meteor trails with lifetime more than 40 seconds, that are out
of the grid and not determined by the algorithm. The sharp decrease
of the distribution below 0.1sec is related with the peculiar properties
of the detection algorithm: the use of the first pulse of the each
sequence (step 1) does not allow us to determine the trail lifetimes
less than the duration of the whole sounding sequence (that has the
order of 0.1 sec). Fig.\ref{fig:FIG5}B shows the distribution of
the meteor trails over the line-of-sight Doppler velocity. From Fig.\ref{fig:FIG5}B
one can see that absolute value of velocity does not exceed 100 m/s.
This confirms correctness of the limits of velocity search. 

Fig.\ref{fig:FIG5}C shows the distribution of the daily number of
meteor trails detected. It shows that usually there are from 150 to
600 meteor trails per day, which corresponds to about 7 to 25 trails
per hour. As the subsequent analysis shows, usually it is enough to
estimate the hourly average neutral velocity. Fig.\ref{fig:FIG5}D
shows the distribution of meteor trails as a function of the radar
range. It can be seen that the main part of the trail observations
is concentrated at ranges less than 300 km. This corresponds well
with the known limitation to the range used for processing meteor
data at SuperDARN radars \cite{JenkinsSDARN_meteorwind}.
Fig.\ref{fig:FIG5}E shows the median velocity, as well as the first
and third quartiles, as a function of local solar time. It can be
seen that the quartiles are much wider than the amplitude of the diurnal
variations of the median velocity, which indicates a high dynamics
of wind changes and the complexity of its description within the framework
of the quasi-isotropic velocity model. Fig.\ref{fig:FIG5}F shows
the median lifetime of the meteor trails, as a function of local solar
time, as well as hourly number of meteors. 
One can see that the lifetime increases in the evening periods,
and decreases in the morning periods. Hourly number of meteors has inverse dynamics.
At present, the lifetime dynamics is
difficult to unambiguously interpret, since the lifetime is a function
of altitude, which can not be detected from our data. 

Fig.\ref{fig:FIG5}F shows the distribution of the number
of echoes detected by the algorithm, depending on the local solar
time. As one can see, the main peak corresponds to the pre-dawn hours,
with the minimum at 15-17 LST. This distribution is similar to the
distributions obtained by the optical methods \cite{LovellMeteorAstronomy}
and other radars \cite{1988JATP...50..703T}.
Dynamics of hourly number of meteors can be explained by 
aspect sensitivity of detected meteor trails.

\subsection{Aspect sensitivity of detected meteor trails}

To verify the correctness of the interpretation of the scattered signal
as underdense meteor trail echo we investigate the aspect character
of the scattering. According to existing point of view, meteor trajectories
during meteor showers is characterized by a certain point on the celestial
sphere - the radiant. Radiants of most showers are accurately calculated
for the day of maximum flow, and are published. Due to lower density the
part of the atmosphere above the meteor combustion height can be considered
as not slowing the meteor and not disturbing its trajectory. So the
meteor trail in the first approximation can be considered as linear
with a known direction, and this direction can be calculated from
the shower radiant. This fact is widely used at standard meteor
radars for estimating shower radiant \cite{CampbellBrown2008144}. The orientation
of the meteor trail irregularity along the burn trajectory and aspect
sensitivity of the scattering at the elongated irregularity lead to
the well-known aspect sensitivity of meteor trail scattering \cite{McKinleyMeteorScience}.

To check the aspect sensitivity of the meteor trails identified by
the algorithm, we chose 13.12.2016 the day of maximum of Geminid
meteor shower with a well known radiant. As a result of processing
EKB radar data there are 877 reliable meteor trails were detected,
with a peak value about 100-120 meteors per hour, which corresponds
well to the forecast data for this shower \cite{IMO_meteor_shower_calendar}.

For each meteor trail detected, we determine its geographic location
from the measured azimuth to the meteor trail and range to it, assuming
its 80km altitude. In the determined geographic point of scattering,
the direction vector was calculated to the radiant of the meteor shower.
Then the aspect angle between the direction to radiant and the direction
from the radar to the scattering point was calculated.

\begin{figure}
\includegraphics[scale=0.07]{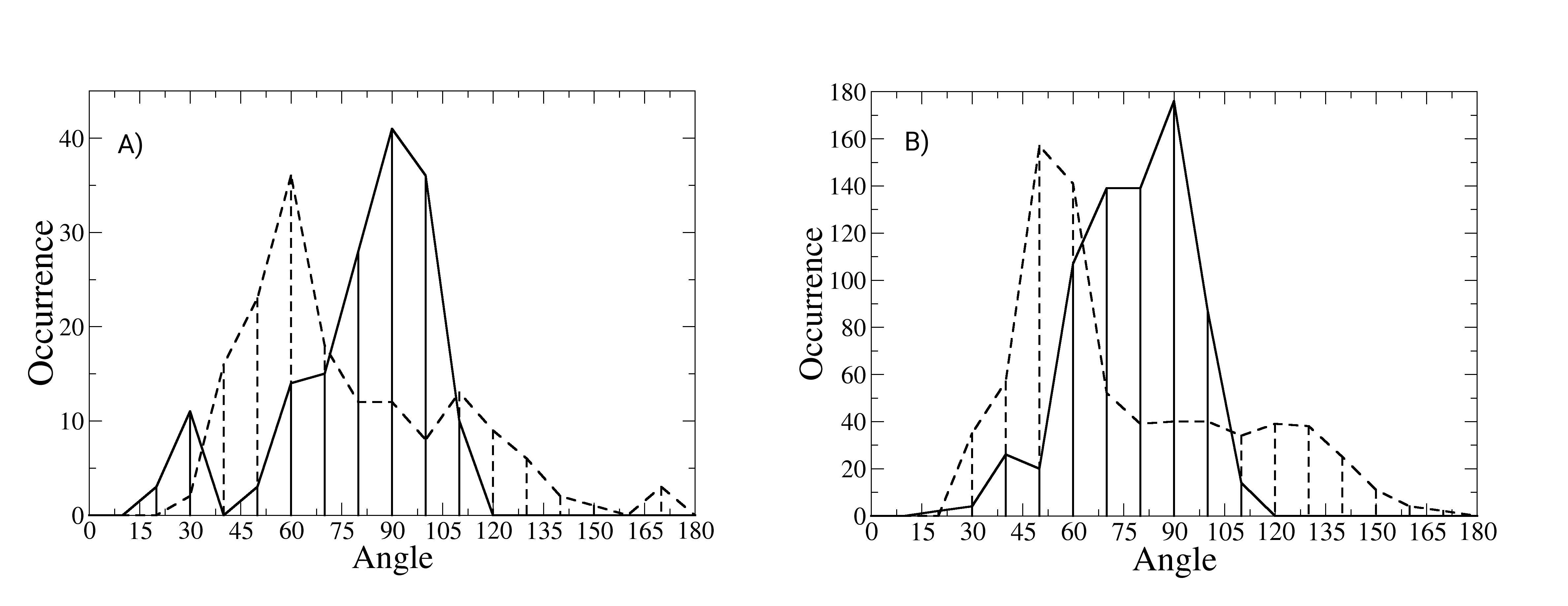}
\caption{Distributions of aspect angles between the direction to the meteor
shower radiant from scattering point and direction to the scattering
point from the radar. A) - 1st channel, B) - 2nd channel. The solid
line is the distribution under approximation of the scattering in
the main antenna lobe, the dashed line is the distribution under approximation
of the scattering in the back lobe.}
\label{fig:FIG6}
\end{figure}

Since the EKB radar did not have the ability to measure the elevation
angle to the scatter during the experiment, and the antenna array
has a significant back lobe, it is possible that the observation of
some meteor trails occurs in the back lobe. Therefore, while estimating
the aspect angle the calculations were made in two cases - in the
case with meteor trail is in the back lobe of the antenna pattern, and
in the case when meteor trail is in the main lobe of antenna pattern.
Fig.\ref{fig:FIG6} shows the distributions of aspect angles
for these two approximations, and for both radar channels used in
the experiment: the solid line corresponds to the scattering in the
main lobe of the antenna pattern, the dashed line corresponds to the
scattering in the back lobe of the antenna pattern. From Fig.\ref{fig:FIG6}
one can see that under the assumption of main lobe scattering the
scattering is aspect sensitive one and most of the meteor trails are
observed when the line of sight is nearly perpendicular to the direction
to the radiant of the meteor shower. This proves that most of the
scattered signals detected by the algorithm are indeed scattering
by meteor trails, and the radar data can be used for estimating neutral
wind.

\section{Estimating the neutral wind}

The main geophysical parameters determined from the meteor trail echo
are the diffusion coefficient, determined from the echo lifetime \cite{Jones_and_Jones1990JATP...52..185J}
and the neutral wind velocity, determined from the Doppler frequency
shift. In this paper we investigate only estimating the neutral wind
velocity.

\subsection{Using weighted least-squares method for estimation of neutral wind
velocity}

Since the radar field-of-view is quite wide, it is possible to formulate
the problem of estimating the full vector of the neutral wind velocity
under the assumption of wind isotropy within the radar field of view.
Under this assumption it is possible to determine the total vector
of the neutral wind velocity by the weighted least-squares method.
The weight function $W(R_{i})$ will be discussed later. The use of
weight function is necessary because of the absence of interferometric
(elevation) measurements at EKB radar. Due to this the combustion
height $h_{0}$ is fixed during calculations. At the same time, a
significant error in determining the horizontal component of Doppler
velocity may exist at small distances.

In the weighted least-squares method, the average wind is determined
from the experimental data, providing the minimum of the residual
functional:

\begin{equation}
\Omega(V_{x},V_{y},V_{z})=\sum_{i}\left\{ W(R_{i})\left(V_{d,i}+V_{x}k_{x,i}+V_{y}k_{y,i}+V_{z}k_{z,i}\right)^{2}\right\} =min\label{eq:VelFunc_3D}
\end{equation}

here $(V_{x},V_{y},V_{z})$ are the projections of the neutral wind
velocity in the local coordinate system of the i-th meteor trail,
where $(X,Y)$ plane is locally parallel to the Earth's surface in
the spherical Earth approximation (X axis is directed to the North,
Y axis is directed to the West, Z axis is directed vertically upwards);
$k_{x,i},k_{y,i},k_{z,y}$ are the components of the unit vector of
the line of sight direction from the radar to the scatter point recalculated
in the scatter local coordinate system; $R_{i}$ is the distance from
the radar to the scatter, $i$ is the index number of the meteor under
study.

The problem of determining the components of a vector by the method
(\ref{eq:VelFunc_3D}) is a simple analytical problem that reduces
to solving a system of three linear equations. The only problem that
needs to be solved is the correct definition of the weight function
$W(R_{i})$. Let's define it.

According to the principles of weighted least square method, for a
normal error distribution the correct weight function $W(R_{i})$
should be inversely proportional to the dispersion of the corresponding
summand in (\ref{eq:VelFunc_3D}), caused by errors:

\begin{equation}
W(R_{i})=\sigma_{V}^{-2}(R_{i})\label{eq:WeightF}
\end{equation}

Thus, to determine the weight $W(R_{i})$ we should estimate the variance
of the velocity as a function of the distance to the scatterer.

\begin{figure}
\includegraphics[scale=0.7]{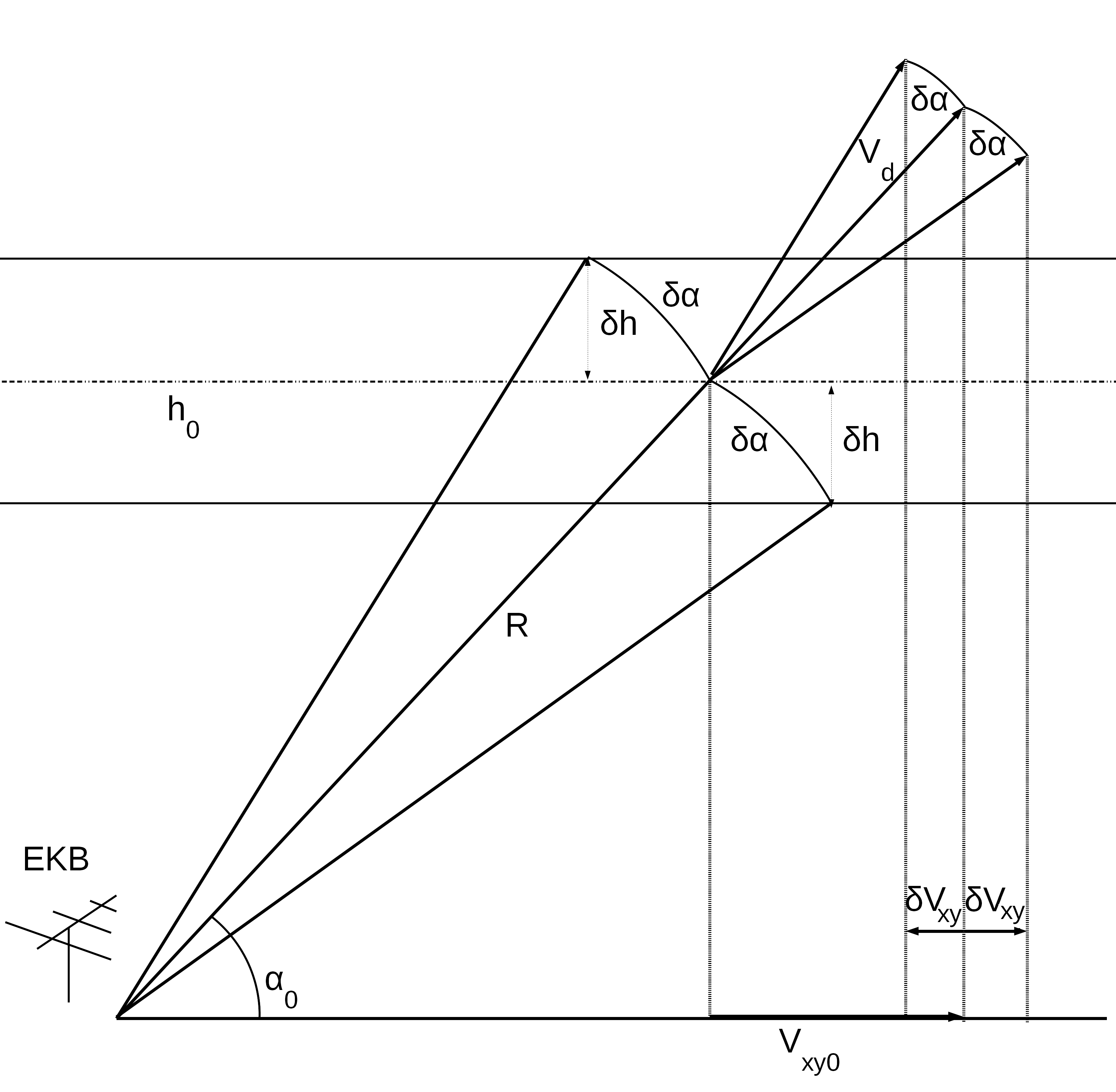}
\caption{Geometry for the calculation of $W(R_{i})$.}
\label{fig:FIG7}
\end{figure}

To estimate the variance of $\sigma_{V}^{-2}(R_{i})$, assume that
the meteor's combustion height has a normal distribution with a mathematical
expectation of $h_{0}$ and a standard deviation of $\sigma_{h}$.
The geometry for the calculations is shown in Fig.\ref{fig:FIG7}.
The random $i$-th meteor produces the trail at a certain height $h_{0}+\delta h_{i}$,
corresponding to the elevation angle $\alpha_{0}(R)+\delta\alpha_{i}$,
where $R$ is the distance to the track.

The projection of the line-of-sight Doppler velocity $V_{d}$ to the
horizontal plane $(X,Y)$ is:

\begin{equation}
V_{xy0}+\delta V_{xy,i}=V_{d}cos(\alpha_{0}(R_{i})+\delta\alpha_{i})\label{eq:V_alpha_rel}
\end{equation}

where $V_{xy0}$ is the mathematical expectation of $V_{xy}$, $\delta V_{xy,i}$
is the random deviation of the horizontal velocity projection, $V_{d}$
is the measured line-of-sight Doppler velocity.

In the first approximation (for $|\delta h_{i}|\ll R_{i}$) from the
geometric considerations shown in Fig.\ref{fig:FIG7},
$\delta\alpha_{i}$ can be represented as:

\begin{equation}
\delta\alpha_{i}=\frac{\delta h_{i}}{Rcos(\alpha_{0}(R_{i}))}\label{eq:alpha_h_rel}
\end{equation}

Let's expand the right side of the (\ref{eq:V_alpha_rel}) to the
power series and discard all the terms smaller than the first order
over the$\delta\alpha_{i}$. After consequent substitution the resulting
expression for $\delta\alpha_{i}$ to (\ref{eq:alpha_h_rel}) we obtain
the following relationship between the variations in horizontal speed
and scatterer height:

\begin{equation}
\frac{\delta V_{xy,i}}{V_{d}}=-\frac{\delta h_{i}}{R}tg(\alpha_{0}(R_{i}))\label{eq:V_vs_H}
\end{equation}

Since the variations in speed and height are proportional to each
other, their variances are also proportional to each other:

\begin{equation}
\sigma_{V}^{2}\left(\frac{\delta V_{xy}}{V_{d}}\right)=\left\{ \frac{tg(\alpha_{0}(R))}{R}\right\} ^{2}\sigma_{h}^{2}\left(\delta h\right)\label{eq:DispV_vs_DispH}
\end{equation}

We suppose that the distribution of the combustion heights is constant
within the radar filed of view, so the dispersion of the scatterer
heights $\sigma_{h}^{2}(\delta h)$ is also constant within field
of view. Due to this, it does not affect the result of (\ref{eq:VelFunc_3D})
and is assumed to be 1.

It should be noted that in the case of $\alpha_{0}(R)=0$ (nearly
horizontal observation of meteor scattering), the line of sight is
in the horizontal plane and $V_{d}$ is nearly equal to the horizontal
velocity. So in this case an uncertainty of the combustion height
in the first approximation does not lead to an uncertainty of the
horizontal velocity. Thus, the expression (\ref{eq:DispV_vs_DispH})
does not contradict with qualitative expectations of relation between
elevation angle accuracy and velocity accuracy.

By substituting the expression for the velocity variance (\ref{eq:DispV_vs_DispH})
into the expression (\ref{eq:WeightF}), we obtain the necessary expression
for the weight function (in the first approximation, valid for $R\ll R_{E};h_{0}\ll R$
, where $R_{E}$ is the Earth radius):

\begin{equation}
W(R)=\frac{R^{2}\left(R^{2}-h_{0}^{2}\right)}{h_{0}^{2}}\label{eq:WeightF_exact}
\end{equation}

The obtained equation is very approximate and does not take into account
spherical Earth, the exact distribution of the meteor trails over
the height, and inaccurate at small distances (less than 300km), but
is a first approximation in the weighted least squares method and
can be used in practice. 

\subsection{Verification of the neutral wind estimation technique}

Due to the absence in the radar field of view the active tools of
monitoring neutral wind at heights of 60-100 km, there is no possibility
for now to directly compare the data received by the radar with the
data of other instruments. So we verify of the efficiency of the method
described indirectly and check the validity of the obtained results
qualitatively. 

As already mentioned, the main component of the neutral wind at these
altitudes is the horizontal wind, its vertical component is often
neglected \cite{manning1954_meteor_wind}. In the framework of the
weighted least squares method considered by us before, the average
two-dimensional (horizontal) wind $(V_{x},V_{y})$ can be determined
from experimental data based on the minimum condition for the following
residual:

\begin{equation}
\Omega(V_{x},V_{y})=\sum_{i}\left\{ W(R_{i})\left(V_{d,i}+V_{x}k_{x,i}+V_{y}k_{y,i}\right)^{2}\right\} =min\label{eq:VelFunc_2D}
\end{equation}
Here all the elements were discussed earlier when discussing the reconstruction
technique of three-dimensional neutral wind (\ref{eq:VelFunc_3D},\ref{eq:WeightF_exact}).

We assume that if the estimated neutral winds are nearly horizontal,
and the horizontal wind rose in 2D model (\ref{eq:VelFunc_2D}) differs
insignificantly from the horizontal wind rose in 3D model (\ref{eq:VelFunc_3D}),
then our technique is valid.

To verify the correspondence between the winds in the two models,
the meteor data during the period 23/12/2016-03/07/2017 was analysed.
During this period the algorithm detected 71219 meteor echoes. Based
on the obtained data, the wind velocity vector was calculated in 2D
model (\ref{eq:VelFunc_2D},\ref{eq:WeightF_exact}) and 3D model
(\ref{eq:VelFunc_3D},\ref{eq:WeightF_exact}) of neutral wind. The
calculation was carried out for different heights of combustion: 80
and 100 km (maximal heights of meteors burning at night and day for
detected lifetime (0.1-40sec) \cite{JGR:JGR3117}).
The statistics was made over the full range of dates, including not
only meteor showers, but also meteorically calm days. The accumulation
time necessary for one measurement of the wind velocity vector was
1 hour. For the rejection of very calm hours and the meteor observations
strongly localized over the azimuths, we limited the hours in which
the calculations can be made. Measurement of wind speed at a given
hour was considered correct under the following conditions: the number
of recorded meteors should exceed 10 meteors per hour, the azimuthal
coverage by meteor trail observations in the radar field-of-view should
be at least 2 beams (6-10 degrees) during the hour.

\begin{figure}
\includegraphics[scale=0.04]{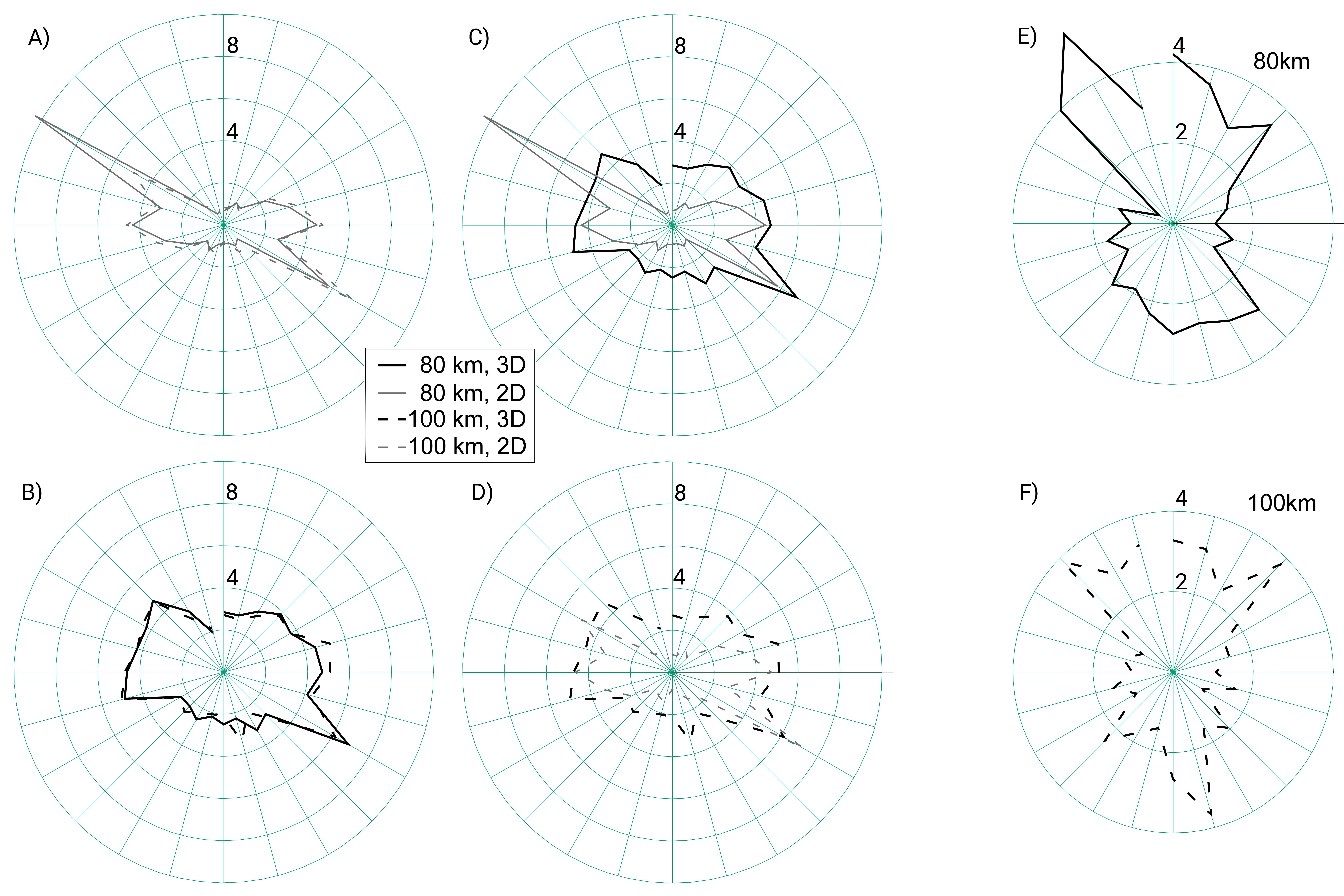}
\caption{Comparison of wind roses obtained for different wind models (3D and 2D) and different meteor trail heights. 
A) - 2D wind model, grey solid line corresponds to 80 km height, grey dashed line corresponds to 100 km height, (in m/s);
B) - 3D wind model, black solid line corresponds to 80 km height, black dashed line corresponds to 100 km height (in m/s); 
C) - comparison of 3d and 2D models at 80km height: black solid line - 3D model, grey solid line - 2D model (in m/s);
D) - comparison of 3d and 2D models at 100km height: black dashed line - 3D model, grey dashed line - 2D model (in m/s);
E) - relationship between rose in 3D model to rose in 2D model for height 80km ;
F) - relationship between rose in 3D model to rose in 2D model for height 100km
 }
\label{fig:FIG8}
\end{figure}

Using obtained wind velocities in the two models, the horizontal wind
roses were constructed (the median value of the horizontal projection
of the wind speed for the selected direction of the wind) with azimuthal
resolution $15^{0}$ . Wind roses are shown in Fig.\ref{fig:FIG8}.
From Fig.\ref{fig:FIG8}A,B one can see that the roses change
slightly with the burning height. From Fig.\ref{fig:FIG8}C,D
one can see that qualitatively the wind roses calculated in 2D
and 3D models are close to each other, and demonstrate the prevalence
of the zonal wind over the meridional, which does not contradict the
existing data \cite{mlt_wind}. At the same time, the median values
of the zonal wind in both models are close, while the meridional wind
within 3D model is 2-3 times higher than the meridional wind in 2D
model(Fig.\ref{fig:FIG8}E,F). This indicates the prevalence of the zonal wind over the vertical
wind, which suggests the validity of the proposed method of wind recovery
from the EKB ISTP SB RAS meteor trail data.

\section{Discussion and conclusion}

In the paper we describe the algorithms for interpreting the signals
scattered on meteor trails and used at EKB ISTP SB RAS radar. The
algorithm is divided into two phases - the detection of underdense
meteor trail echo and the estimation of its parameters. At the first
phase, spatially localized bursts of amplitude above the average signal
level, monotonically decaying in time are detected. In the second
phase, the quadrature components of the scattered signal are approximated
by the exponential model with Doppler shift by the method of least
squares, by performing additional filtering of the found meteor trails
based on the resulting accuracy of phase and amplitude approximation.
In general, the algorithm is similar to the algorithms used for specialized
meteor radars, but when restoring the scattering parameters it approximates
the quadrature components of the received signal by the model instead
of separate approximating the amplitude and the phase of the signal.

To test the algorithm, it was shown that, on the day of the maximum
of the Geminid shower 13.12.2016, the scattering detected by the
algorithm has an aspect sensitive character and correspond to scattering
at irregularities elongated with the direction toward the radiant
of the meteor shower. This confirms that the source of signals detected
by the algorithm is scattering on meteor trails.

In the paper, the solution is presented for reconstructing the vector
of the neutral wind velocity from the obtained data by the weighted
least-squares method under the assumption of the isotropy of the field
of neutral winds in the local coordinate system of the meteors. Based
on geometric considerations and assuming a Gaussian distribution of
the meteor trail heights the weight function was found and the inversion
technique was implemented at EKB radar.

To test the inversion technique the neutral winds were reconstructed
in the models of two-dimensional (horizontal) wind and three-dimensional
wind based on the EKB radar long-term measurements (from 23.12.2016
to 03.07.2017). The roses of the horizontal neutral winds obtained
in both models are compared, their proximity and the expected predominance
of the zonal wind over the meridional one are shown. The comparison
showed that statistically the vertical wind is smaller than the zonal
wind.

The developed algorithms allow processing the scattered signals in
real time. The algorithm for detecting the meteor trails and estimating
their parameters (the Doppler shift and the lifetime), has been put
into regular operation at the EKB ISTP SB RAS radar.

\section*{Acknowledgements}
In the paper the data from EKB ISTP SB RAS radar were used. The work
was supported by FSR program II.12.2.3.

\end{document}